\documentclass[a4paper,11pt]{article}
\usepackage{jcappub} 
\usepackage{amsfonts,amsmath,amssymb,amsthm,bbm,hyperref}
\usepackage{color,psfrag}
\usepackage{graphicx}
\newcommand{\chicrit}{1/(3\sqrt{\kappa})}

\makeatother

\begin{document}


\title{Quantum behavior of the ``Little Sibling'' of the Big Rip induced by a three-form field}

\author[a,b]{Mariam Bouhmadi-L\'opez}
\author[a]{David Brizuela}
\author[a]{I\~naki Garay}

\affiliation[a]{Department of Theoretical Physics, University of the Basque Country
UPV/EHU, P.O. Box 644, 48080 Bilbao, Spain\\}

\affiliation[b]{IKERBASQUE, Basque Foundation for Science, 48011, Bilbao, Spain}

\emailAdd{mariam.bouhmadi@ehu.eus}
\emailAdd{david.brizuela@ehu.eus}
\emailAdd{inaki.garay@ehu.eus}


\abstract{
A canonical quantization \`{a} la Wheeler-DeWitt is performed for a model of three-form fields in a homogeneous and isotropic universe. We start by  carrying out the Hamiltonian formalism of this cosmological model.  We then  apply this formalism to a Little Sibling of the Big Rip (LSBR), an abrupt event milder than a Big Rip and that is known to be generic to several minimally coupled three-form fields for a variety of potentials. We obtain a set of analytical solutions of the Wheeler-DeWitt equation using different analytical approximations and explore the physical consequences of them. It turns out that there are quantum states where the wave function of the universe vanishes, i.e. the DeWitt condition is fulfilled for them. Given that this happens only for some subset of solutions of the Wheeler--DeWitt equation, this points out that the matter inducing the LSBR is equally important in the process as, it has been previously shown, a minimally coupled phantom scalar field feeding classically a LSBR is smoothed at the quantum level, i.e. all the quantum states lead to a vanishing wave function.}



\maketitle

\flushbottom

\section{Introduction}

General relativity predicts the existence of singularities in a wide variety of models, being the Big Bang the most common in a cosmological setting. It turns out that this is not the only possible cosmological singularity in a 
classical universe. In fact, the current cosmological data  \cite{Ade:2015xua,Abbott:2017wau} are compatible with a future doomsday, not predicted by the conventional $\Lambda$CDM model.

One of the mildest possible future doomsdays is what is known as the Little Sibling of the Big Rip (LSBR) \cite{Bouhmadi-Lopez:2014cca,MCP}. It is an abrupt event at an infinite cosmic time, therefore it is not a singularity, 
however all the known structures in our universe would be ripped apart at a finite cosmic time if a LSBR would be  the final stage of our universe  the LSBR \cite{Bouhmadi-Lopez:2014cca}. This cosmic event is much smoother than the Big Rip singularity. When it is reached, the Hubble rate and the scale factor blow up but the cosmic time derivative of the Hubble rate does not. Therefore, the scalar curvature explodes. Most importantly, it has been shown that such a model is compatible with the current acceleration of the universe and it happens whenever the sum of the pressure of dark energy and its energy density is constant and negative no matter how small is such a constant \cite{Bouhmadi-Lopez:2014cca}. Consequently, the tiniest deviation from a cosmological constant could unavoidably induce a LSBR. It has been equally shown that a LSBR is ubiquitous in models with three-form matter fields \cite{Morais:2016bev,Bouhmadi-Lopez:2016dzw}. 

Matter of the kind of $p$-forms has been invoked in cosmology already for a while \cite{Germani:2009iq,Koivisto:2009sd,Germani:2009gg}; being 0-forms, i.e. scalar fields, and three-forms the simplest one to be embedded in cosmology given their straightforward compatibility with a homogeneous and isotropic space-time. On this work, we will focus on three-form fields which have been extensively used on the recent years as a mean to describe the primordial inflationaty period or the late-time speed up of the universe \cite{Koivisto:2009fb,Koivisto:2009ew,DeFelice:2012jt,DeFelice:2012wy,Mulryne:2012ax,Kumar:2014oka,Kumar:2016tdn}, not surprisingly given that back in the eighties they were used as a mean to explain the cosmological constant problem \cite{Duff:1980qv}. We will review on the next section how a three-form for some potentials (with quite a broad shape) can induce easily a LSBR \cite{Morais:2016bev,Bouhmadi-Lopez:2016dzw}.

Whenever the issue of singularities or abrupt events emerge, it is hoped that a quantum theory of gravity could erase or at least appease them \cite{clausbook}. Unfortunately, so far there is not a complete and fully consistent  theory of quantum gravity nor of quantum cosmology despite the efforts and multiple candidates in the market. Here, we will apply one of the oldest and most conservative approaches based on a canonical quantization \`{a} la Wheeler--DeWitt (WDW) and see if there are states, i.e. solutions of such an equation, that could lead towards a possible resolution of the LSBR. 

The paper is outlined as follows, on the next section we review the emergence of the LSBR in the framework of three-forms. On section \ref{sec:hamiltonian}, we will construct the classical Hamiltonian and obtain the corresponding WDW equation for three-forms, as far as we know this is done here for the first time. Then in section \ref{sec:WdWLSBR}, we  solve such an equation  by using suitable approximations. We find  different solutions of the wave function, we show that there is a whole set of solutions that can satisfy the DeWitt condition and therefore could hint towards the quantum resolution of the LSBR from a quantum perspective. Yet as we stress on the conclusion, in section \ref{sec:conclusions}, the DeWitt condition is not fulfilled for all the wave functions and while a LSBR is fully avoided when induced by a phantom scalar field this is not the case when it is feed by a three-form. This is in striking difference with a LSBR induced by a phantom scalar field where, as has been shown previously, all the solutions of the WDW equation vanish. We include as well an appendix where we proof the validity of the approximations used in section \ref{sec:WdWLSBR}.

\section{Three-forms and the raise of the LSBR: a review}
\label{review}

Let us consider a three-form field $A_{\mu\nu\rho}$ weakly coupled to gravity, ruled by the action \cite{Koivisto:2009sd,Koivisto:2009fb,Koivisto:2009ew}, 
\begin{equation}
 S=\int d^4x \sqrt{-g}\left[\frac{R}{12 \kappa}-\frac{1}{48}F^{\mu\nu\rho\sigma}F_{\mu\nu\rho\sigma}-V(A^{\mu\nu\rho}A_{\mu\nu\rho})\right],
\label{action}
\end{equation}
with
\begin{equation}
 F_{\mu\nu\rho\sigma}=4\nabla_{[\mu}A_{\nu\rho\sigma]}=\nabla_{\mu}A_{\nu\rho\sigma}-\nabla_{\sigma}A_{\mu\nu\rho}+\nabla_{\rho}A_{\sigma\mu\nu}-\nabla_{\nu}A_{\rho\sigma\mu},
\end{equation}
$R$ is the Ricci scalar of our space-time, and $\kappa=4\pi G/(3c^4)$. From now on, we will restrict to positive potentials in the action (\ref{action}). Please, notice that we are interested on the asymptotic behavior of the universe where baryonic and dark matter are subdominant as compared with dark energy played by the 3-form field, and thus they are not considered here.

It can be shown that the equation of motion for this three-form reads 
\begin{align}
	\label{general_eq_motion}
	\nabla_{\sigma}{F^{\sigma}}_{\mu\nu\rho}
		-12\frac{\partial \,V}{\partial 	\left(A^2\right)} A_{\mu\nu\rho}=0
		\,,
\end{align}
where, for simplicity, we have defined $A^2\equiv A^{\alpha\beta\gamma}A_{\alpha\beta\gamma}$.

In the case of a spatially flat Friedmann-Robertson-Walker-Lema\^{i}tre (FLRW) metric
\begin{equation}
ds^2=-dt^2+a^2(t)d\vec{x}^2,
\end{equation}
we can write the non-zero components of the three-form field in the following way~\cite{Germani:2009iq}:
\begin{equation}
\label{A_ijk_comp}
A_{ijk}=\chi(t)\sqrt{h}\epsilon_{ijk}=\chi(t)a^3\epsilon_{ijk},
\end{equation}
where $\epsilon_{ijk}$ stands for the 3-dimensional Levi-Civita symbol and $\chi(t)$ is a scalar quantity related to the three-form. 
Therefore, it can be shown that the non-zero components of the strength tensor reads
\begin{align}
	\label{F_0ijk_comp}
	F_{0ijk} = a^{3}(t)\left[\dot\chi(t)+3H(t)\chi(t)\right]\epsilon_{ijk}
	\,,
\end{align}
where $ H(t)=\dot{a}/a$ is the Hubble parameter.
Finally, making use of the Friedmann equation~\cite{Koivisto:2009ew}
\begin{align}
	\label{NFriedm-1-1}
	 H^{2}=2\kappa\left[
		\frac{1}{2}\left(\dot{\chi}+3 H\chi\right)^{2}
		+V
	\right]
	\,,
\end{align}
the equation of motion for the field $\chi(t)$ can be written as 
\begin{align}
	\label{chi_eq_motion}
	\ddot{\chi}+3 H\dot{\chi}+\left[1 - \left(\frac{\chi}{\chi_\mathrm{c}}\right)^2\right]V_{,\chi}=0
	\,,
\end{align}
where $\chi_\mathrm{c}=\chicrit$,  and $V_{,\chi}$ stands for the derivative of the potential, $V$, with respect to $\chi$. Note that this equation looks pretty much similar to that of a minimally coupled scalar field with an effective potential, $V^{\mathrm{eff}}$, such that $V^{\mathrm{eff}}_{,\chi}=(1- (\chi/\chi_\mathrm{c})^2 )V_{,\chi}$.
Therefore, the former equation of motion implies that a homogeneous and isotropic three-form admits as stationary solutions the points where (i) $V_{,\chi}=0$ or (ii) the points $\chi = \pm\chi_\mathrm{c}$. In what refers the fixed points $\pm\chi_\mathrm{c}$, it can be shown that 
\begin{align}
	\label{second_derivative_Veff}
	V^\textrm{eff}_{,\chi\chi}\left(\pm\chi_ \textrm{c}\right) 
	= - 2\frac{V_{,\chi}\left(\pm\chi_ \textrm{c}\right)}{\pm\chi_{\textrm{c}}} 
	= -4 V_{,\chi^2}\left(\pm\chi_ \textrm{c}\right)
	\,.
\end{align}
Thus, whenever the potential is a decreasing function of $\chi^2$ at $\pm\chi_\mathrm{c}$, those points could correspond to stable fixed points towards which the three-form  evolves naturally. In fact, this is the case as has been shown thoroughly in Ref.~\cite{Morais:2016bev,Bouhmadi-Lopez:2016dzw}.

In addition, the Raychaudhuri equation reads
\begin{align}
	\label{NHdot-1-1}
	\dot{ H}=-3\kappa\chi V_{,\chi}=-6\kappa\chi^2V_{,\chi^2}
	\,,
\end{align}
Consequently, even though the kinetic energy density of the three-form is positive, this kind of matter can induce a super-inflationary era, i.e. an increase of the Hubble parameter, whenever $V_{,\chi^2}$ is negative. This is precisely the case for the fixed points $\pm\chi_\mathrm{c}$ if they are attractors, given that 
\begin{equation}
\dot{H}_{(\chi\rightarrow\pm\chi_\mathrm{c})}=
-\frac23\,V_{,\chi^2}(\pm\chi_\mathrm{c}).
\end{equation}
Therefore, and shown in Ref. \cite{Morais:2016bev} by dynamical system analysis,
\begin{align}
	  H^2_{(\chi\rightarrow\pm\chi_\mathrm{c})} \sim -\frac43 \,V_{,\chi^2}(\pm\chi_\mathrm{c})\,\ln(a)
	\,.
\label{asymptoticH}
\end{align}
This implies that in an expanding universe, the Hubble parameter will blow up as the scale factor diverges if these fixed points are attractors.

In summary, what we have shown is that if a potential at the fixed points 
$\pm\chi_\mathrm{c}$ is positive and a decreasing function of $\chi^2$, then the Hubble parameter blows up while its cosmic time derivative is finite. It can be equally shown that this happens at an infinite cosmic time (cf. Eq.~(\ref{asymptoticH}) and Ref.~\cite{Morais:2016bev}).
Consequently, if the following conditions are meet: 
 positiveness and  non-vanishment of the potential at $\pm\chi_\mathrm{c}$ with  $V_{,\chi^2}$ negative at the same points, the three-form values $\pm\chi_\mathrm{c}$ correspond to stable fixed points describing  LSBR abrupt events. Such abrupt events do not correspond to  dark energy singularities given that they take place at an infinite cosmic time but share pretty much the rest of the characteristics of dark energy singularities for that all the known structures in the universe would be destroyed when  the LSBR is approached. 

\begin{figure}
\begin{center}
       \includegraphics[width=0.7\textwidth]{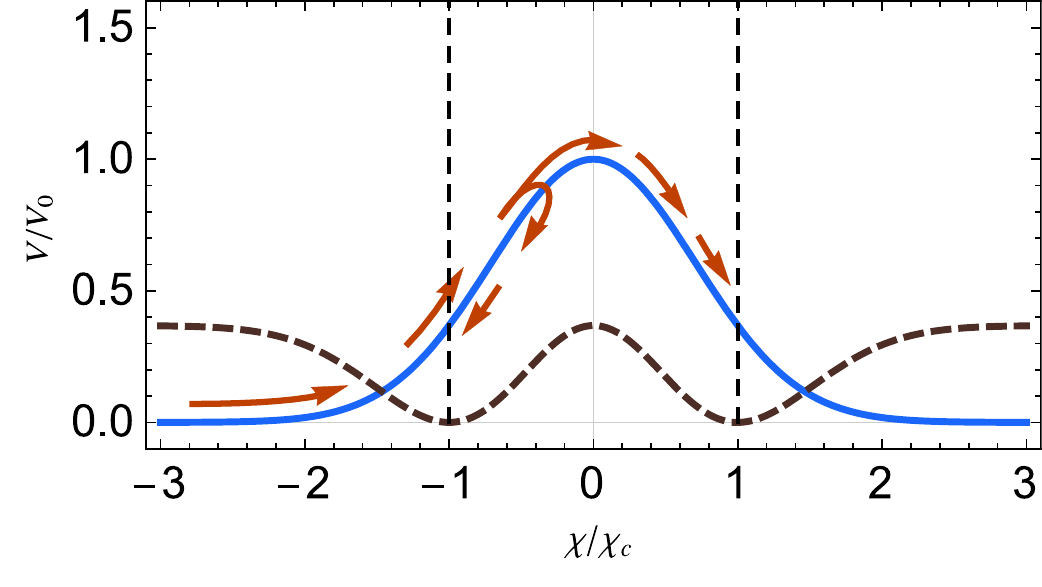}
\end{center}
       \caption{This plot shows the dynamics of a three-form for a Gaussian potential (blue continuous 
       curve). As proven in Refs.~\cite{Koivisto:2009ew,Koivisto:2009fb} for positive-valued $V$, as it is the case of a Gaussian potential, the Friedmann equation implies $(\dot\chi+3H\chi)^2<(3H\chi_\mathrm{c})^2$. Therefore, for an expanding universe and values $|\chi|>\chi_\mathrm{c}$ implies that $\dot\chi\chi<0$, i.e., independently of the shape of the potential, the field $\chi$ evolves monotonically towards the region $[-\chi_\mathrm{c},\chi_\mathrm{c}]$ in which it remains trapped. The brown dashed plot represents $V^\textrm{eff}$ up to a shift constant.}
    \label{potentialdynamics}
\end{figure}

We will further impose that the  square speed of sound of the three-forms, $c_{s}^{2}=\frac{\chi V_{,\chi\chi}}{V_{,\chi}}$, is positive at the fixed points to avoid any potential classical instability at the perturbative level. It can be shown that a Gaussian potential fulfills  all these conditions and the dynamics of the three-form can be seen schematically in Fig.\ref{potentialdynamics}. A careful and exhaustive analysis of a three-form with a Gaussian potential playing the role of dark energy (in presence or absence of interaction with dark matter) can be found in Ref.~\cite{Morais:2016bev,Bouhmadi-Lopez:2016dzw}.

In order to quantize this model and seek the possible avoidance of the LSBR, we will first get the classical Hamiltonian describing this system for an arbitrary potential and next obtain the corresponding Wheeler-DeWitt equation also in full generality. This is carried out on the next section.

\section{Hamiltonian formalism and the Wheeler-DeWitt equation}
\label{sec:hamiltonian}

\subsection{The classical Hamiltonian}

In order to obtain the classical Hamiltonian, we will first get the symmetry reduced Lagrangian of the system.  Using the relations given in Eqs. (\ref{A_ijk_comp}) and (\ref{F_0ijk_comp}), the following equalities hold:
 \begin{eqnarray}
&&A^{\mu\nu\rho}A_{\mu\nu\rho}=6\chi^2\,\\
&& F^{\mu\nu\rho\sigma}F_{\mu\nu\rho\sigma}=-24(\dot{\chi}+3 H\chi)^2=-24\left[\frac{1}{a^3 N}\frac{d}{dt}\left(a^3 \chi\right)\right]^2.
\label{equations}
\end{eqnarray}
Given equation (\ref{equations}), it is natural to define a new pseudo-scalar field $\phi=a^3\chi$. This last term is the one that appears in
the action and this transformation makes the Lagrangian to be diagonal in the sense that is composed by
a linear combination of quadratic time derivatives. This property will make the Hamiltonian diagonal
in $(a,\phi)$ variables and, thus, is the most natural to proceed to the quantization.

Using the former equations, we obtain the following expression for the action of the three-form:
\begin{equation}
S_A=\int dt\, {\cal V} N\sqrt{h}\left(\frac{1}{2N^2}\frac{\dot{\phi}^2}{a^6}-V
\right),
\end{equation}
${\cal V}$ being the volume, in principle infinite, of our homogeneous spatial section. From now on, we will assume spatially flat section, given (i) the observations are consistent with spatially flat sections and (ii) the spatial curvature term is sub-dominant when the LSBR is approached. 

On the other hand, the Einstein-Hilbert action for this case takes the form 
\begin{equation}
S_{EH}=-\frac{1}{2\kappa}\int dt\, {\cal V}\frac{a\dot{a}^2}{N}.
\end{equation}
Therefore, the total action of our system will be
\begin{equation}
S=S_{A}+S_{EH}=\int dt\, {\cal V} N \left(\frac{-a\dot{a}^2}{2\kappa N^2}+\frac{\dot{\phi}^2}{2a^3N^2}-a^3V
\right)
.
\end{equation}
It is straightforward to see that the fiducial volume ${\cal V}$ can be absorbed in our fundamental variables by
performing the rescaling $a\rightarrow{\cal V}^{1/3}a$ and $\phi\rightarrow{\cal V}\phi$. In particular note that
this transformation leaves invariant the argument of the potential.

At this point, we define the conjugate moments of our fundamental variables in the usual way,
\begin{eqnarray}
 p_a=\frac{\delta L}{\delta \dot{a}}=-\frac{a\dot{a}}{\kappa N},\\
 p_\phi=\frac{\delta L}{\delta \dot{\phi}}=\frac{\dot{\phi}}{a^3 N}.
\end{eqnarray}
Finally, after a Legendre transformation, the Hamiltonian of our system is obtained:
\begin{equation}
\mathcal{H}=N\left(-\frac{\kappa}{2a}p_a^2+\frac{a^3}{2}p_\phi^2+a^3V
\right).
\label{Hamiltonian}
\end{equation}
This is the Hamiltonian that will be used to perform the quantization in the next section.
But, before that, let us briefly comment the form the Hamiltonian would take if one uses
the variable $\chi$ instead of $\phi$. In fact, it is easy to obtain it by considering the canonical transformation $(a, p_a,\phi, p_{\phi})\rightarrow(a,\tilde{p}_a,\chi,p_{\chi})$,
\begin{equation}
\phi=a^3\chi,\qquad
p_{\phi}=p_\chi/a^3,\qquad
p_a=\tilde{p}_a-\frac{3}{a}\chi p_\chi,
\end{equation}
which implies a change in the momentum of the scale factor $a$.
In this way, the Hamiltonian would take the form,
\begin{equation}
\mathcal{H}=N\left(-\frac{\kappa}{2a}(\tilde{p}_a-\frac{3}{a}\chi p_\chi)^2+\frac{1}{2a^3}p_\chi^2+a^3V
\right),
\end{equation}
which, as commented above, is non-diagonal since it mixes moments $p_\chi$ and $\tilde p_a$. For simplicity, from now on, we will restrict our analysis to the Hamiltonian given in Eq.~(\ref{Hamiltonian}).

\subsection{Wheeler-DeWitt quantization}
\label{sec:WdW}

We can write the classical Hamiltonian (\ref{Hamiltonian}) as:
\begin{equation}
\mathcal{H}=N\left(\frac{1}{2}G^{AB}p_A p_B+a^3V\left(6(a^{-3}\phi)^2\right)\right),
\end{equation}
with $A$ and $B$ indices referring to $a$ or $\phi$ and the mini-superspace metric given by
\begin{equation}
G^{AB}=\left(
\begin{array}{cc}
-\frac{\kappa}{a}& 0\\
0& a^3
\end{array}
\right),
\qquad
G_{AB}=\left(
\begin{array}{cc}
-\frac{a}{\kappa}& 0\\
0& \frac{1}{a^3}
\end{array}
\right).
\end{equation}
With the usual prescription, we transform the classical dynamical variables into quantum operators by means of the Laplace-Beltrami operator:
\begin{equation}
G^{AB}p_A p_B\rightarrow -\frac{\hbar^2}{\sqrt{-G}}\partial_A(\sqrt{-G}G^{AB}\partial_B)\,,
\end{equation}
where $G$ is the determinant of $G_{AB}$.

Therefore, we obtain the equation $\hat{H}\psi=0$, that is the Wheeler-DeWitt equation,
\begin{equation}
\left(
\hbar^2 \kappa\partial^2_\beta-\hbar^2\partial^2_\phi+2V
\right)
\psi (\beta,\phi)=0,
\label{WdWeq}
\end{equation}
where we have introduced the new variable $\beta=a^3/3$ related with the volume of the spatial section of the FLRW metric that we are considering. Notice that the kinetic operator related to the geometry, i.e. $\beta$, and matter, i.e. $\phi$, have opposite sign. This is natural as, although the three-form is mimicking a phantom behavior, it is not truly phantom matter.

\section{Quantum cosmology of the LSBR with a three-form field as matter source}
\label{sec:WdWLSBR}

As we have mentioned in section \ref{review}, the following Gaussian potential gives rise to the scenario where the so-called LSBR is found:
\begin{equation}
V=V_0 e^{-\lambda^2 \chi^2}
=V_0 e^{-9\lambda^2 \phi^2/\beta^2}\,,
\end{equation}
with $\lambda^2=\kappa/2\sigma^2$, and $\sigma$ the dimensionless width of the Gaussian potential. The WDW equation is given now by:
\begin{equation}
\left(\hbar^2 \kappa\partial^2_\beta-\hbar^2\partial^2_\phi+2V_0 e^{-9\lambda^2 \phi^2/\beta^2}\right)\psi(\beta,\phi)=0.
\label{WdWequation}
\end{equation}
Our aim is to analyze the asymptotic behavior  of the solutions  of this equation for large $\beta$ and check whether they are damped and thus avoid quantum mechanically the abrupt event LSBR obtained in the classical picture. In order to do it, we will make use of two different and complementary approximations. On the one hand, a semiclassical approximation,
where it will be assumed a nearly classical behavior of our variables. In particular, as already commented above,
classically $\chi$ tends to a constant value $\chi_c$. Thus, an expansion of the potential will be performed up to
linear order in $(\chi-\chi_c)$. On the other hand, a particular set of potentials, those with large width, will be considered
more explicitly by performing a quadratic expansion in the parameter $\lambda$ of the potential.

\subsection{Semiclassical approximation}

\begin{figure}
\begin{center}
       \includegraphics[width=0.6\textwidth]{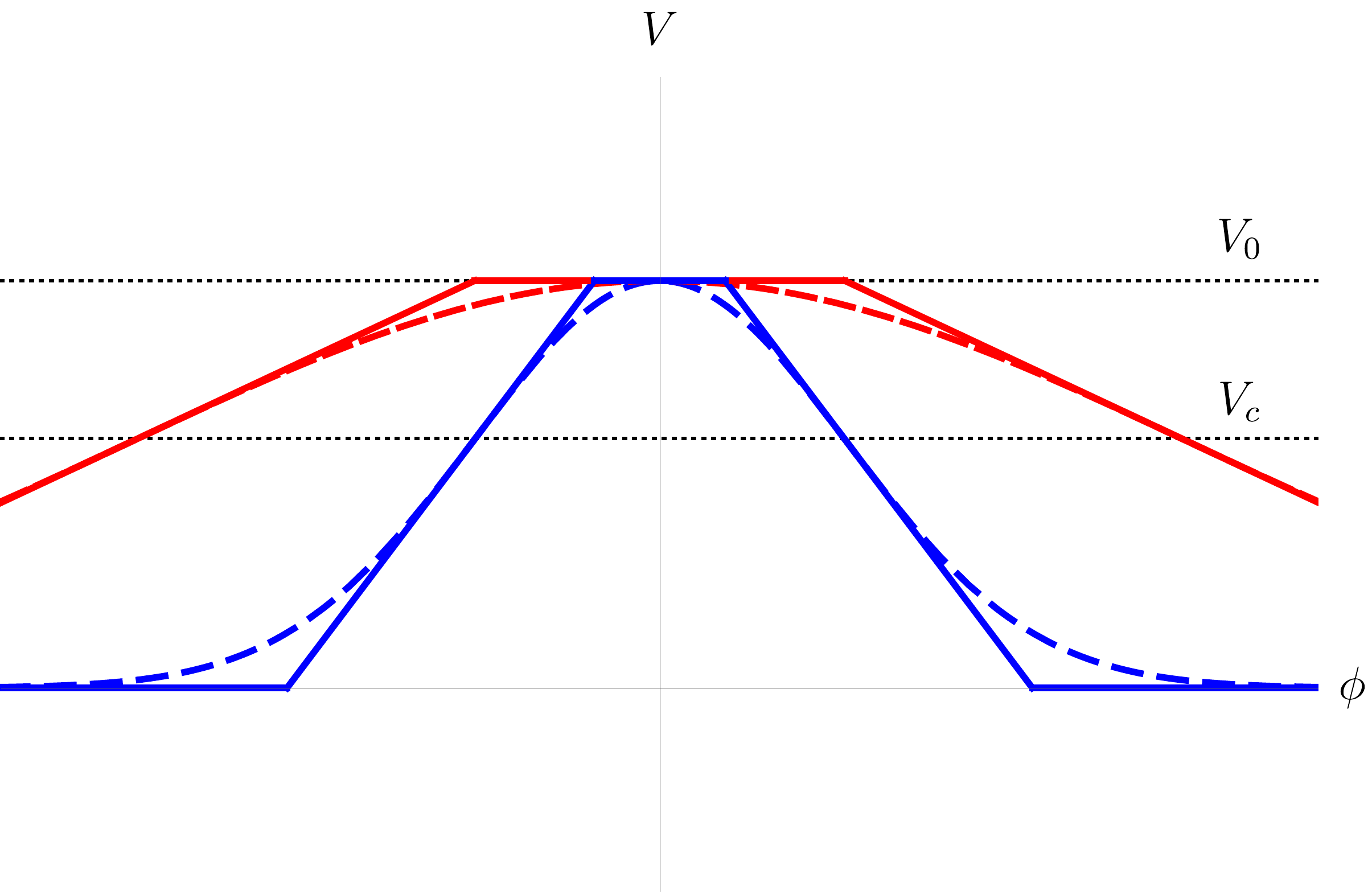}
\end{center}
       \caption{Two Gaussians are shown for early times (short dashed line) and for later times (long dashed line). The continuous lines show the approximate potentials that will be used to solve the WDW equation. Finally, the dotted lines stand for the maximum $V_0$ value of the potential and its asymptotic value $V_c$.}
    \label{linearapproximation}
\end{figure}

By expanding the potential around its value in $\chi_c$ we get the following approximate form 
\begin{equation}\label{linearpotential}
V=V_c-\gamma\left(\frac{\phi}{3\beta}-\chi_c\right),
\end{equation}
where orders higher than quadratic in $(\chi-\chi_c)$ have been dropped. In this expression, the following definitions have been made:
\begin{equation}
V_c=V_0e^{-\lambda^2\chi_c^2},\qquad\gamma=2\lambda^2\chi_c V_c. 
\end{equation}
Note that this potential for $\phi$ is time-dependent, and the Gaussian is in fact broaden out as $\beta$ increases, as it tends to a constant $V_c$. In Fig.~\ref{linearapproximation} we show two Gaussians which correspond to different values of $\beta$, with their corresponding linear approximation tangent at the point $\chi_c$. As can be seen clearly in the figure, a good approximation for these Gaussians are the poligonal functions drawn with continuous lines. Thus, we will solve the Wheeler-DeWitt equation for constant and linear potentials in the following subsections. The former case
can be analitically solved whereas, for the latter one,
a Born--Oppenheimer approximation will be applied. We recall that this approximation was used in cosmology for the first time in Ref.~\cite{BOClaus} and it implies that one part of the wave function evolves adiabatically with respect to one of the degrees of freedom of the system. Once the solutions
are at hand, in order to find a complete solution for the system, one should just impose continuity of the function and its first derivative at the matching points. We will not perform such a computation though, since our main interest lies on the asymptotic ($\beta\rightarrow\infty$) behavior of the solutions.

\subsubsection{Solution for constant potential}

In the central and in the asymptotic (for large $\phi$) regions  of the Gaussian the potential can be assumed to be constant, $V_0$ or $0$ respectively (see Fig.~\ref{linearapproximation}). In such a case, all the solutions that decay with $\beta$  as $\phi$ follows an almost classical behavior $\phi\approx 3\chi_c\beta$, and therefore obey the DeWitt condition, are written as:
\begin{eqnarray}
\psi(\beta,\phi)&=&\int_0^{V_k}e^{-\sqrt{2 (V_k-E)/\hbar^2}|\phi|} \left(c_1(E)e^{-i\sqrt{\frac{2E}{\hbar^2\kappa}}\beta} + c_2(E)e^{i\sqrt{\frac{2E}{\hbar^2\kappa}}\beta} \right)dE\nonumber\\
&&+\int_{-\infty}^0 c_3(E)\exp\left(-\sqrt{2 (V_k-E)/\hbar^2}|\phi|-\sqrt{\frac{2|E|}{\hbar^2\kappa}}\beta\right)dE\,,
\label{const_pot}
\end{eqnarray}
where $V_k$ must be chosen either $0$ or $V_0$ depending the region of interest, and $c_1(E)$, $c_2(E)$, $c_3(E)$ are arbitrary functions related with the initial conditions. Note that, in this case, all the decaying solutions correspond to classically forbidden modes ($E<V_k$). The rest of solutions are either free oscillating modes (for $E>V_k$) or exponentially divergent solutions (for $E<0$).

\subsubsection{Linear order}

In the regions where the poligonal potential is not constant, it is described by the linear function given in \eqref{linearpotential}. For solving the equation with such an approximate linear potential, we will consider a Born-Oppenheimer approximation.
For that, we will assume a form of the wave function given by $\psi(\beta,\phi)=\sum_n b_n(\beta) \varphi_n(\beta,\phi)$. Plugging it
into the WDW equation, one gets,
\begin{equation}
\hbar^2 (\kappa b_n\partial^2_{\beta}\varphi_n
+2\kappa b'_n\partial_\beta \varphi_n  +\kappa\varphi_n b''_n-b_n\partial^2_\phi \varphi_n)
+2 V b_n \varphi_n=0\,.
\label{wdwbo_linear}
\end{equation}

In the Born-Oppenheimer approximation one neglects the first and second terms inside the parenthesis of that equation, i.e. we are assuming that $\varphi_n$ evolve adiabatically with respect to $\beta$.
This can be safely done as long as the following inequalities are obeyed:
 \begin{equation}
\left|\frac{b_n\partial^2_\beta \varphi_n}{\varphi_n b''_n}\right|\ll 1, \qquad
\left|\frac{b'_n\partial_\beta \varphi_n}{\varphi_n b''_n}\right|\ll 1\,.
\label{bo_conditions}
\end{equation}
The validity for this approximation will be discussed in appendix \ref{semiclassicalBO}.

Therefore, in this approximation, the above equation can be decoupled and one obtains two differential equations. A time-independent Schr\"odinger equation for the functions $\varphi$,
\begin{equation}
-\frac{\hbar^2}{2}\partial^2_\phi \varphi_n+V\varphi_n=E_n\varphi_n,
\end{equation}
and a harmonic oscillator kind of equation for the coefficients $b_n$,
\begin{equation}
\frac{\hbar^2\kappa}{2}b_n''+E_n b_n=0\,.
\label{wkb}
\end{equation}
Note that $E_n=E_n(\beta)$, and although we can not give an explicit expression for it, we can obtain an approximate solution of Eq.~(\ref{wkb}) using the WKB method:
\begin{equation}
b_n(\beta)\sim \left(\frac{2E_n(\beta)}{\hbar^2\kappa}\right)^{\!\!-1/4}
\left[c_1 \exp\left(
i\int\!\sqrt{\frac{2E_n(\beta)}{\hbar^2\kappa}}d\beta
\right)\!+
c_2 \exp\left(
-i\int\!\sqrt{\frac{2E_n(\beta)}{\hbar^2\kappa}}d\beta
\right)
\right],
\end{equation}
with constants $c_1$ and $c_2$. As expected, oscillatory solutions for positive energies and exponential solutions for negative energies are obtained. In the latter case, we will have to choose only the decaying solutions to ensure the appropriate decay of the wave function. We remember at this regard that large values of $\beta$ corresponds classically to the LSBR.

For the linear potential \eqref{linearpotential}, the solution of the Schr\"odinger equation can be written in terms of a linear combination of first and second kind Airy functions:
\begin{equation}
 \varphi_n(\beta,\phi)=c_1 Ai(u)+c_2 Bi(u),
\end{equation}
with the argument
\begin{equation}
u=-\left(\frac{3\sqrt{2}\beta}{\hbar\gamma}\right)^{2/3}\left(E_n-\gamma\chi_c-V_c+\gamma\frac{\phi}{3\beta}\right)\,.
\end{equation}
Note that, in the semiclassical regime, for large values of $\beta$, the ratio $\phi/(3\beta)$ tends to $\chi_c$.
Thus, for large $\beta$, the argument of the Airy functions takes the form,
\begin{equation}
 u\sim-\left(\frac{3\sqrt{2}\beta}{\hbar\gamma}\right)^{2/3}\left(E_n-V_c\right).
\end{equation}

The behavior of the Airy functions for large $\beta$ (that is, large $|u|$) in the case $E_n>V_c$ is given as (cf. Eqs. 10.4.60 and 10.4.64 of Ref.~\cite{Abram}, respectively),
\begin{eqnarray}
&&Ai(u)=Ai(-|u|)\sim \frac{1}{|u|^{1/4}\sqrt{\pi}}\sin\left(\frac{2}{3}|u|^{3/2}+\frac{\pi}{4}\right)\xrightarrow{\,\beta\rightarrow\infty\,}0,\\
&&Bi(u)=Bi(-|u|)\sim \frac{1}{|u|^{1/4}\sqrt{\pi}}\cos\left(\frac{2}{3}|u|^{3/2}+\frac{\pi}{4}\right)\xrightarrow{\,\beta\rightarrow\infty\,}0,
\end{eqnarray}
whereas for the case $E_n<V_c$ one gets (cf. Eqs. 10.4.61 and 10.4.63 of Ref.~\cite{Abram}, respectively),
\begin{eqnarray}
&&Ai(u)=Ai(|u|)\sim \frac{1}{2|u|^{1/4}\sqrt{\pi}}\exp\left(-\frac{2}{3}|u|^{3/2}\right)\xrightarrow{\,\beta\rightarrow\infty\,}0,\\
&&Bi(u)=Bi(|u|)\sim \frac{1}{|u|^{1/4}\sqrt{\pi}}\exp\left(\frac{2}{3}|u|^{3/2}\right)\xrightarrow{\,\beta\rightarrow\infty\,}\infty.
\end{eqnarray}
In order to impose the DeWitt condition one should select the decaying solutions for large $\beta$:
\begin{eqnarray}
E_n>V_c&\Rightarrow & \varphi_n=c_1 Ai(u)+c_2 Bi(u)\,,\\
E_n<V_c&\Rightarrow & \varphi_n=c_3 Ai(u)\,.
\end{eqnarray}
Notice that in the energy range $E_n>V_c$ all solutions asymptotically decay and would thus avoid the LSBR assuming that the vanishing of the wave function would imply zero probability of finding the system in such a state. Nevertheless, in the case $E_n<V_c$ only the Airy functions of the first kind are suitable decaying solutions.

Therefore, aside from getting the whole set of solutions of the WDW equations under the assumed Born--Oppenheimer approximation, we have as well proven that there is a whole bunch of solutions that fulfill the DeWitt condition when the LSBR region is approached, and thus would provide a quantum regularization of the LSBR.

\subsection{Quadratic approximation}

We will study a further approximation to the Gaussian potential. In the following, we assume that potential can be expanded as 
\begin{equation}
V=V_0e^{-9\lambda^2(\phi/\beta)^2}\sim V_0(1-9\lambda^2(\phi/\beta)^2)\,,
\end{equation}
which is valid as long as  $9\lambda^2\phi^2/\beta^2\ll 1$, that is good for wide Gaussians $9\kappa\phi^2/2\beta^2\ll\sigma^2$. This approximations has important physical consequences as the wider is the Gaussian the larger is the interval over which the square of the speed of sound of the three-form is positive. We remind at this regard that for a Gaussian potential, close to the LSBR the square of the speed of sound of the three-form is always positive (cf. Sec.~\ref{review}).

In order to solve the Wheeler DeWitt equation (\ref{WdWequation}) we  consider the Born-Oppenheimer (BO) approximation for this equation, so we look for solutions of the form $\psi(\beta,\phi)=\sum_n g_n(\beta,\phi)f_n(\phi)$. In this case, the WDW equation is
\begin{equation}
\hbar^2 \kappa f_n\partial^2_\beta g_n
-\hbar^2(f_n\partial^2_\phi g_n+2 f'_n\partial_\phi g_n  +g_n f''_n)
+2V_0(1-9\lambda^2(\phi/\beta)^2)f_ng_n=0\,,
\label{WdW_BO}
\end{equation}
where the following conditions have to be satisfied
\begin{equation}
\left|\frac{f_n\partial^2_\phi g_n}{g_n f''_n}\right|\ll 1, \qquad
\left|\frac{f'_n\partial_\phi g_n}{g_n f''_n}\right|\ll 1\,.
\label{conditions}
\end{equation} 
We will check the validity of this approximation in the appendix \ref{quadraticBO}. Let us now discuss the possible solutions to this equation.

\begin{figure}
\begin{center}
            \includegraphics[width=0.6\textwidth]{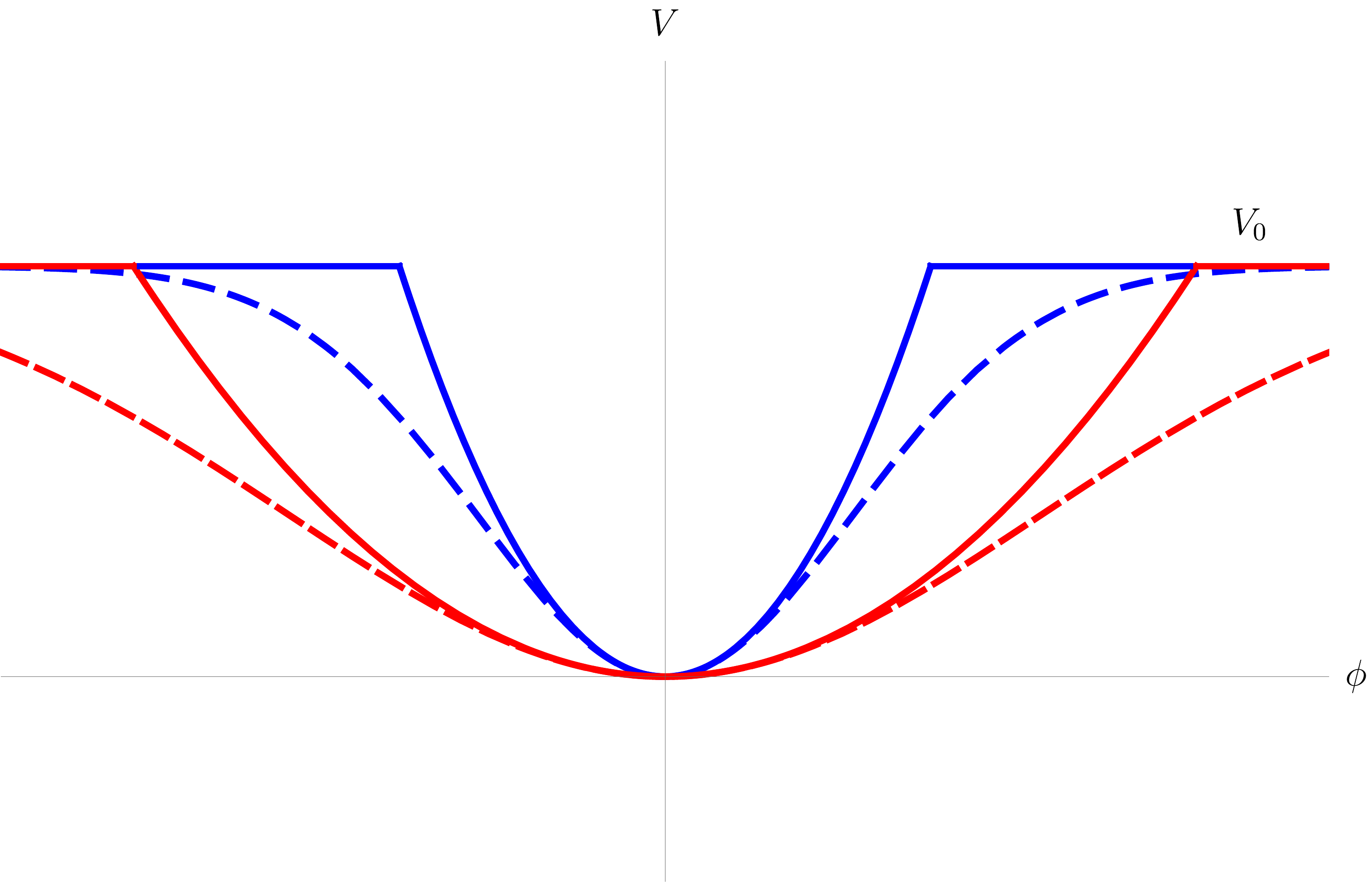}
\end{center}
        \caption{Two Gaussians are shown for early times (short dashed line) and for later times (long dashed line). The continuous lines show the approximate potentials that will be used to solve the WDW equation.}
    \label{figparabola}
\end{figure}

In this approximation, the WDW equation (\ref{WdW_BO}) leads, therefore, to the following two differential equations:
\begin{eqnarray}
&& -\frac{\hbar^2 \kappa}{2}\partial^2_\beta g_n-V_0(1-9\lambda^2(\phi/\beta)^2)g_n=(E_n-V_0)g_n\,,
\label{eq_g}
\\
&& \frac{\hbar^2}{2} f''_n+(E_n(\phi)-V_0)f_n=0\,,
\end{eqnarray}
where the separation constant has been chosen as $(E_n-V_0)$ in order to have the same origin of energies as in
the previous section. Equation \eqref{eq_g} corresponds to a Schr\"odinger equation with an effective potential $9V_0\lambda^2(\phi/\beta)^2$. In fact, in order to properly approximate the Gaussian potential, one should just consider this parabolic form until it takes a value $V_0$ and a constant potential $V_0$ elsewhere (see Fig.~\ref{figparabola}).

The solutions of the equation for $g_n$ are written in terms of the Bessel functions of the first and second kind as:
\begin{equation}
g_n=\sqrt{\beta}\left(
c_{1}J_\nu(\sqrt{\epsilon_n}\beta)+c_{2}Y_\nu(\sqrt{\epsilon_n}\beta)
\right)\,,
\end{equation}
with
$$
\nu=\frac{1}{2}\sqrt{1+72V_0\lambda^2\phi^2/(\hbar^2\kappa)}\,,
\qquad
\epsilon_n=\frac{2E_n}{\hbar^2\kappa}
$$
and $c_{1},c_{2}$ integration constants.

For the functions $f_n$ we can obtain an approximate solution of the equation as it was done in the previous section, using the WKB method
\begin{eqnarray}
f_n(\phi)&\sim & \left(\frac{2(E_n(\phi)-V_0)}{\hbar^2}\right)^{-1/4}
\left[c_1 \exp\left(
i\int\sqrt{\frac{2(E_n(\phi)-V_0)}{\hbar^2}}d\phi
\right)\right.\nonumber\\
&+ &\left.c_2 \exp\left(
-i\int\sqrt{\frac{2(E_n(\phi)-V_0))}{\hbar^2}}d\phi
\right)
\right]\,,
\end{eqnarray}
with constants $c_1$ and $c_2$. We will obtain oscillatory or exponential solutions depending on the signs of $E_n(\phi)$. As before, in the exponential case one could choose the negative exponential in order to obtain the appropriate decay of the wave function.

The asymptotics of the Bessel functions of the first and second kind is (cf. Eqs.~9.2.1 and 9.2.2 of Ref.~\cite{Abram}):
\begin{equation}
J_\nu(z)\sim\sqrt{\frac{2}{\pi z}}\cos\left(z-\frac{\nu\pi}{2}-\frac{\pi}{4}\right)\,,
\qquad
Y_\nu(z)\sim\sqrt{\frac{2}{\pi z}}\sin\left(z-\frac{\nu\pi}{2}-\frac{\pi}{4}\right)\,,
\end{equation}
with $z\in \mathbb{C}$, $|\text{arg}(z)|<\pi$, $|z|\rightarrow\infty$ and $\nu\in\mathbb{R}$.
This gives the following behavior,
\begin{equation}
g_n\sim \exp\left[\pm i\left(
\sqrt{2E_n}\,\beta-\frac{\nu\pi}{2}-\frac{\pi}{4}
\right)\right].
\label{asymp_behav}
\end{equation}

Therefore, in our case (where $\nu$ is real and the argument of the Bessel functions may be positive or purely imaginary depending on the sign of the energies $E_n$), taking into account the previous asymptotic expansions, it is easy to check that only in the $E_n<0$ case is possible to construct decaying solutions by considering the following combination:
\begin{equation}
g_n(\beta,\phi)= c_{3}\sqrt{\beta} \left(J_\nu(\sqrt{\epsilon_n}\beta)+ i\, Y_\nu(\sqrt{\epsilon_n}\beta)\right)\,.
\label{asymp_g_n}
\end{equation}
In this way, we conclude that for values $E_n>V_0$ both $f_n$ and $g_n$ tend to freely oscillating modes as the system approaches the LSBR. On the other hand, in the region $0<E_n<V_0$, $f_n$ tend to exponentially growing and decaying modes whereas the $g_n$ are oscillating functions. Finally, for $E_n<0$  both tend to real exponential solutions. In summary, for energies below $V_0$ it is possible to construct decaying solutions that would obey the DeWitt condition.

\section{Conclusions}
\label{sec:conclusions}

One of the biggest and most challenging issues that theoretical physics is facing is the true nature of dark energy or whatever is  currently fuelling the accelerated expansion of our universe. While it might be possible that our universe will be inflating forever as a boring $\Lambda$CDM universe, where so far no convincing solution to the true nature of the cosmological constant is available at hand, it might be that dark energy is an evolving component and in particular of a phantom-like nature \cite{Ade:2015xua}. Within this context, it is well known that different future doomsdays are observationally possible \cite{MCP}, being what is known as the LSBR one of the possibilities. This abrupt event which is not a true cosmological singularity as it happens at an infinite cosmic time, though the scalar curvature blows up when approaching it, share some feature with the Big Rip singularity on the sense that at a finite cosmic time from now all the known bounded structure would be destroyed \cite{Bouhmadi-Lopez:2014cca}. It has been also shown recently that a three-form field induce naturally a LSBR in a homogeneous and isotropic universe. It is therefore natural to address the quantization of the system as a mean to look for a potential resolution of the LSBR.

In order to tackle the quantization of a three-form in a FLRW universe we have obtained its classical Hamiltonian (\ref{Hamiltonian}) after defining proper variables that diagonalize the two momentum of the system. As can be seen the quadratic three-form  momentum enters with a positive sign into the Hamiltonian, as it is expected, because although a three-form can induce naturally a super-acceleration phase, i.e. a phantom-like behavior, it is not phantom in its essence. In fact, the obtained Hamiltonian is equivalent to that of a canonical scalar field with a time dependent potential.  This has drastic consequences on the obtained WDW equation~(\ref{WdWeq}), by means of the Laplace--Beltrami operator, which has a hyperbolic nature rather than an elliptic nature if we where dealing with true phantom matter (cf. for example Ref.~\cite{Albarran:2015cda}).

We have considered the WDW equation with a Gaussian potential that may lead to a LSBR behavior \citep{Morais:2016bev}. In particular, the WDW equation has been solved for two different approximations for the potential. On the one hand a semiclassical approximation where we have considered an expansion of the potential around the classical asymptotic value of the field $\chi$. And, on the other hand, a quadratic approximation valid for the particular case of wide Gaussians has been developed. 

In both approximations we have concluded that, as expected, all modes with $E<0$ are either exponentially damped  or amplified (in the two variables considered $\beta$ and $\phi$) because they correspond to classically forbidden solutions. For modes with $0<E<V_0$ one obtains the exponentially increasing and decreasing solutions for one of the variables and oscillatory behavior for the other one. Finally, modes with $E>V_0$ are freely oscillating solutions (in both variables $\beta$ and $\phi$). Therefore, for energies of the modes below $V_0$ one could impose the DeWitt condition and construct solutions that would avoid the abrupt event LSBR. This is not the case though for modes with energy over $V_0$. In any case, one should keep in mind that this is based in a probabilistic interpretation which, although extensively used, would only be valid after constructing a  Hilbert space. 

Before concluding we would like also to comment on the differences between the quantum cosmology of phantom scalar field that induces a LSBR and the quantum cosmology of a three-form inducing the same LSBR. While at the classical regime both type of matter induces the same kind of Hubble expansion asymptotically this is not the case at the quantum level. Indeed, while all the wave functions solution of the WDW equation vanishes for a phantom minimally couple scalar field, see Ref.~\cite{Albarran:2015cda}, this is not the case for a three-form field. Therefore, the DeWitt condition is better fulfilled for a phantom scalar field than for a three-form field. 


\acknowledgments
We thank Manuel Kr\"amer and Jo\~ao Morais for fruitful discussions.
The authors acknowledge financial support from
project FIS2017-85076-P (MINECO/AEI/FEDER, UE), and
Basque Government Grant No.~IT956-16.
This paper is based upon work from COST action CA15117 (CANTATA),
supported by COST (European Cooperation in Science and Technology).
The work of MBL is supported by the Basque Foundation of Science Ikerbasque.


\appendix

\section{Validity of the Born-Oppenheimer approximations}

\subsection{Semiclassical approximation}\label{semiclassicalBO}

Let us study now the validity regime of the Born-Oppenheimer approximation developed previously \eqref{bo_conditions}.
First we will assume that the eigenvalues $E_n(\beta)$ vary very slowly, that is, we will assume $E_n'(\beta)\sim 0$. This assumption is sensible in the relevant limit for large values of $\beta$. Therefore, we obtain for the functions $b_n(\beta)$ the following asymptotic behavior:
\begin{equation}
b'_n\sim \sqrt{\frac{2E}{\hbar^2\kappa}}b_n\,,
\qquad b''_n\sim \frac{2E}{\hbar^2\kappa}b_n\,.
\end{equation}

Then, considering the asymptotics of the Airy functions, we can write the solutions to the Schr\"odinger equation as
\begin{equation}
\varphi_n(\beta,\phi)\sim\exp(w_n(\beta,\chi_c))\,,\qquad 
w_n(\beta,\chi_c)=-\frac{2\sqrt{2}}{\hbar\gamma}(E-V_c)^{3/2}\beta-\frac{1}{6}\log\beta;
\end{equation}
and we can easily compute the derivatives
\begin{eqnarray}
&& \frac{\partial_\beta \varphi_n}{\varphi_n}= \partial_\beta w_n\sim\frac{2\sqrt{2}}{\hbar\gamma}(E-V_c)^{3/2}+\mathcal{O}(1/\beta)\,,
\\
&& \frac{\partial^2_\beta \varphi_n}{\varphi_n}= (\partial_\beta w_n)^2+\partial^2_\beta w_n\sim \left(\frac{2\sqrt{2}}{\hbar\gamma}\right)^2(E-V_c)^{3}+\mathcal{O}(1/\beta)\,.
\end{eqnarray}

The conditions (\ref{bo_conditions}) are then reduced to the following expression:
\begin{equation}
\sigma^4\ll\kappa\chi_c^2\frac{E V_c^2}{|E-V_c|^3}.
\end{equation}
The dimensionless quantity $\kappa\chi_c^2\sim 1$ 
(as stated in the seminal paper \cite{Morais:2016bev} where the Gaussian potential was invoked to study the LSBR abrupt event. Please see also the definition of just after Eq.~(\ref{chi_eq_motion})). Therefore, our solutions will be valid for
\begin{equation}
\sigma^4\ll\frac{E V_c^2}{|E-V_c|^3}\,.
\end{equation}
Therefore, for wide Gaussians, we could only consider those modes with energies $E\sim V_c$ in order the BO approximation to be valid. Nevertheless, for narrow potentials, a large range of energies of the modes could contribute to the non-divergent solution.

\subsection{Quadratic approximation}\label{quadraticBO}

We will study now the conditions that we have to impose on our solutions $\psi_n=g_n f_n$ in order to fulfill the BO approximation (equations \ref{conditions}).

The validity of this approximation can be explored using expansions with respect to $\kappa$ \cite{BouhmadiLopez:2009pu}. On the one hand, for the functions $f_n(\phi)$ we obtain (considering as before that $E'_n(\phi)\sim 0$):
\begin{equation}
f'_n\sim \frac{\sqrt{|E_n-V_0|}}{\hbar}f_n\,,
\qquad f''_n\sim \frac{|E_n-V_0|}{\hbar^2}f_n\,.
\end{equation}
On the other hand, for the functions $g_n$ we will study their asymptotic behavior for large $\beta$ (\ref{asymp_behav}). Therefore, considering the functions $g_n$ written as
\begin{equation}
g_n\sim \exp(h_n(\beta,\phi))\,;\qquad h_n(\beta,\phi)=\pm i\left(
\sqrt{2E_n}\,\beta-\frac{\nu\pi}{2}-\frac{\pi}{4}
\right)\,.
\end{equation}
we obtain the following results:
\begin{eqnarray}
&& \frac{\partial_\phi g_n}{g_n}= \partial_\phi h_n\sim \sqrt{\frac{V_0}{\hbar^2\kappa}}\lambda +\mathcal{O}(\kappa^{1/2})\,,
\\
&& \frac{\partial^2_\phi g_n}{g_n}= (\partial_\phi h_n)^2+\partial^2_\phi h_n\sim\sqrt{\frac{\hbar^2\kappa}{V_0}}\frac{1}{\phi^3\lambda}+\mathcal{O}(\kappa^{3/2})\,.
\end{eqnarray}

Finally, using the equations (\ref{conditions}), we obtain that
\begin{equation}
\sqrt{\frac{V_0}{|E_n-V_0| \kappa}}\lambda\ll 1\,, \qquad
\frac{\hbar^4\kappa}{V_0 |E_n-V_0| \lambda \phi^3}\ll 1\,.
\end{equation}

The second of these equations is automatically satisfied in the desired limit ($\beta\rightarrow\infty$), given that we are considering that the field $\phi$ diverges in that limit. The first equation gives us the relation:
\begin{equation}
\frac{V_0}{\sigma^2}\ll|E_n-V_0|\,.
\end{equation}
That is, the wider the Gaussian of the potential, the wider the range of the modes that we can consider in order to construct regular solutions with the appropriate behavior at the LSBR.



\end{document}